\documentclass[prb,aps,twocolumn]{revtex4}
\usepackage{amssymb}
\usepackage{color}
\usepackage{graphicx}
\usepackage{bbm}
\usepackage{amsfonts}
\usepackage{mathrsfs}
\usepackage{amsmath}

\begin{document}

\title{Statistical fluxes and the Curie-Weiss metal state}
\author{Kai Wu,$^{1}$ Zheng-Yu Weng,$^{1}$ and Jan Zaanen$^{2}$}
\affiliation{$^{1}$Institute for Advanced Study, Tsinghua University, Beijing, 100084, China\\
$^{2}$Instituut Lorentz for Theoretical Physics, Leiden University, Leiden, The Netherlands}

\begin{abstract}
We predict a new state of matter in the triangular t-J model in a high doping regime. Due to the altered role of quantum statistics the spins are `localized' in statistical Landau orbits, while the charge carriers form a Bose metal that feels the spins through random gauge fields. In contrast to the Fermi-liquid state, this state naturally exhibits a Curie-Weiss susceptibility, large thermopower, and linear-temperature resistivity, explaining the physics of \textrm{Na}$_{x}$\textrm{CoO}$_{2}$ at $x>0.5.$ A `smoking gun' prediction for neutron scattering is presented.
\end{abstract}

\date{\today}
\maketitle

\emph{Introduction.---}Among the great challenges in quantum matter physics is the question whether Mottness -- the drastic change of Hilbert space structure due to strong local interactions -- might form the condition for non Fermi liquid states of fermionic matter to occur \cite{Anderson,Overbosch}. Among others, this is the point of departure of Anderson's resonating valence bond (RVB) proposal for high-$T_{c}$ superconductivity \cite{anderson1}. At the electron densities in the proximity of the half-filled Mott insulator the Mott projections are most evident and it is much less evident whether they matter at all at very high(or dilute) electron densities. Resting on a representation \cite{Overbosch,reviewZYW} that makes explicit the altered nature of the quantum statistics due to Mottness we demonstrate here the existence of an
internally consistent, stable saddle point at high dopings on a triangular lattice. This is a spin-charge separated state where the doublons (carrying charge) and spinons (carrying spin) are coded as hard core bosons that communicate with each other via statistical fluxes. The surprise is in the spin sector: the statistical gauge fields act like a uniform magnetic field causing Landau quantization of the spinon states. The spinons are localized in Landau orbits of the lowest Landau level (LLL) at low temperatures, with the effect that these behave as free Curie-Weiss spins with large entropy. The doublons get in turn scattered by the random gauge fluxes associated with the localized spinons, similarly as in gauge glass models.

This might sound far fetched. However, the microscopy matches that of the highly overdoped $\mathrm{Na}_{x}\mathrm{CoO}_{2}$ system \cite{Teraski,Thermopower,Phase} with $x>0.5$. A very strange metal is formed in
this cobaltate, characterized by a high density of free Curie-Weiss spins that `appear out of the blue' while it is a rather good hole-type thermoelectric material \cite{mhlee} given its large thermopower combined with a relatively low resistivity. We will show here that the magnetic-(Fig. 1), thermoelectric- (Fig. 2) and resistive (Fig. 3) properties of the cobaltate are consistent with the state described in the previous paragraph. Moreover, we present a `smoking gun' prediction that can be straightforwardly tested by experiment: the spin excitation spectrum should carry the fingerprints of the `spontaneous' Landau quantization (Fig. 4). We notice that earlier attempts \cite{PALee,shastry,Kuroki} to explain the Curie-Weiss metal behavior of the cobaltates relied on the assumption of a very small bare bandwidth, while in our theory the small itineracy scale is emergent.

Similar to the layered structure in the high-$T_{c}$ cuprates, the CoO$_{2}$ layer is believed to play an essential role in determining the low-energy physics in the \textrm{Na}$_{x}$\textrm{CoO}$_{2}$ compound. With electrons doping introduced by \textrm{Na}, the electron hopping at the partially filled \textrm{t}$_{2g}$ orbitals of the \textrm{Co}$^{+3}$ ions and spin correlations between \textrm{Co}$^{+4}$ ions in a CoO$_{2}$ layer may be minimally described by the t-J model \cite{RVB1,RVB2,RVB3} on a triangular lattice with hopping integral $t<0$ \cite{arpes,arpes2,zwang}. Here the
Hilbert space is constrained by $\sum_{\sigma }c_{i\sigma }^{\dagger }c_{i\sigma }\geq 1$, i.e., each lattice site is either singly occupied by an electron (\textrm{Co}$^{+4})$ or doubly occupied by electrons (\textrm{Co}$^{+3})$ without allowing the empty site (\textrm{Co}$^{+5})$. What we will be interested in is the highly overdoped regime of this model, where the RVB \cite{RVB1,RVB2,RVB3} correlations induced by the superexchange coupling disappear and the hopping processes of both charge and spin become dominant.

\emph{Sign structure.---}To identify the new saddle point, we have to rely on a particular representation that is superficially reminiscent of the standard slave bosons but is actually quite different. This `phase string'
representation \cite{Overbosch,reviewZYW} is making explicit the non Fermi-Dirac nature of the quantum statistics in doped Mott insulators. According to this formalism, the electron operator $c_{i\sigma }$ can be fully `bosonized'\ by $c_{i\sigma }^{\dagger }=d_{i}^{\dagger }b_{i-\sigma}e^{-i\Theta _{i\sigma }^{\text{\textrm{string}}}}$, where the doublon\ and spinon\ creation operators, $d_{i}^{\dagger }$ and $b_{i\sigma }^{\dagger }$, are both \emph{bosonic}, which satisfy an equality (Mott) constraint $n_{l}^{d}+\sum_{\sigma }n_{l\sigma }^{b}=1$ with $n_{l}^{d}$ and $n_{l\sigma}^{b}$ denoting the doublon and spinon occupation numbers, respectively. Here the fermionic commutations of the electron operators are ensured by the
topological phase $\Theta _{i\sigma }^{\text{\textrm{string}}}\equiv \lbrack\Phi _{i}^{s}+\sigma \Phi _{i}^{d}]/2$, with $\Phi _{i}^{s}\equiv\sum_{l\neq i}\theta _{i}(l)\sum_{\sigma }\sigma n_{l\sigma }^{b}$ and $\Phi_{i}^{d}\equiv \sum_{l\neq i}\theta _{i}(l)\left( 1-n_{l}^{d}\right) $,where $\theta _{i}(l)={Im}$ $(z_{l}-z_{i})$ ($z_{i}$ is the complex coordinate of site $i$). In this representation, the hopping and superexchange terms of the t-J model become
\begin{eqnarray}
H_{t} &=&t\sum_{\langle ij\rangle \sigma }\hat{D}_{ji}\hat{B}_{ij}^{\sigma}+H.c.  \label{Ht} \\
H_{J} &=&-{\frac{J}{2}}\sum_{\langle ij\rangle\sigma}\hat{B}_{ij}^{\sigma}\hat{B}_{ji}^{-\sigma}-{\frac{J}{2}}\sum_{\langle ij\rangle\sigma}n_{i\sigma }^{b}n_{j-\sigma }^{b},  \label{Hj}
\end{eqnarray}
in terms of $\hat{D}_{ij}\equiv e^{iA_{ij}^{s}}d_{i}^{\dagger }d_{j}$ and $\hat{B}_{ij}^{\sigma }\equiv e^{-i\sigma A_{ij}^{d}}b_{i\sigma }^{\dagger}b_{j\sigma }.$ In this way, the remnant sign structure after Mott projection has been made explicit while it is precisely represented by the topological link variables, $A_{ij}^{s}$ and $A_{ij}^{d}$ --- Obviously
there would be no `sign problem' should $A_{ij}^{s}=A_{ij}^{d}=0$\ in such a fully bosonized model with $t<0$. Here $A_{ij}^{s}$ and $A_{ij}^{d}$ satisfy
\begin{eqnarray}
\sum_{\langle ij\rangle \in \partial S}A_{ij}^{s} &=&\pi \sum_{l\in S}\sum_{\sigma }\sigma n_{l\sigma }^{b}  \label{as} \\
\sum_{\langle ij\rangle \in \partial S}A_{ij}^{d} &=&\pi \sum_{l\in S}\left(1-n_{l}^{d}\right)  \label{ad}
\end{eqnarray}
where $\partial S$ denotes the boundary of an area $S$.

\emph{Electron fractionalization saddle-point.---}The precise sign structure identified above will be crucial in constructing the following saddle-point which respects the gauge invariance associated with $A_{ij}^{s}$ and $A_{ij}^{d}$, together with time-reversal and spin rotational symmetries. Since the RVB pairing is irrelevant at high doping, the gauge-invariant $\hat{D}_{ij}$ and $\hat{B}_{ij}^{\sigma }$ in Eqs. (\ref{Ht}) and (\ref{Hj}) will be natural order parameters, resulting in an effective $H_{\mathrm{eff}}=H_{d}+H_{s}$:
\begin{eqnarray}
H_{d} &=&-t_{d}\sum_{\langle ij\rangle }e^{-ia_{ij}}\left(e^{iA_{ij}^{s}}d_{i}^{\dagger }d_{j}\right) +H.c.  \label{hd} \\
H_{s} &=&-t_{s}\sum_{\langle ij\rangle \sigma }e^{-ia_{ij}}\left(e^{-i\sigma A_{ij}^{d}}b_{i\sigma }^{\dagger }b_{j\sigma }\right) +H.c. \label{hs}
\end{eqnarray}
where $t_{d}=-tB>0$, $t_{s}=-tD+JB/4>0$. Here $a_{ij}$ represents the U(1) gauge fluctuations around the saddle-point: $\hat{D}_{ij}\simeq De^{ia_{ij}}$ and $\hat{B}_{ij}^{\sigma }\simeq (B/2)e^{ia_{ij}}$. It can be shown \cite{long version} that the \emph{transverse} fluctuations of $a_{ij}$ will get suppressed \cite{semion} to ensure the stability of the saddle-point and the flux binding conditions Eqs. (\ref{as}) and (\ref{ad}), implying spin-charge separation. Note that the fractionalized state as governed by $H_{\mathrm{eff}}$ at the gauge fixing $a_{ij}=0$ is still an intrinsic gauge model
involving the mutual Chern-Simons gauge fields, $A_{ij}^{s}$ and $A_{ij}^{d}$.

\begin{figure}[tb]
\includegraphics[angle=0,width=\columnwidth]{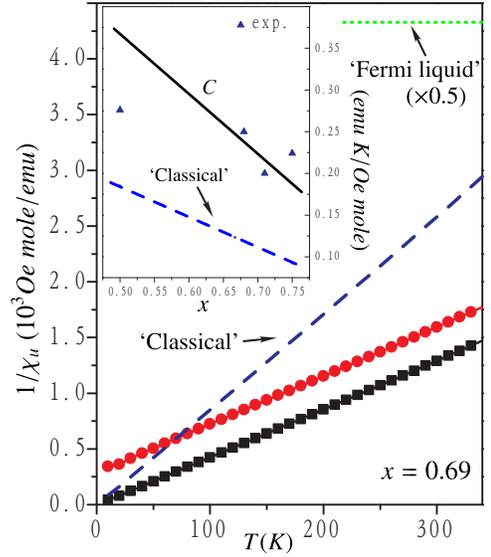}
\caption{The spin part of the present electron fractionalization state
exhibits a Curie $T$-dependent uniform susceptibility $\protect \chi _{u}=C/T$
(full squares) at $x=0.69$. The Weiss term in $\protect \chi _{u}$ appears as
an RPA correction (full circles). The Fermi liquid (dotted) and `classical
limit'\ (dashed) results are also shown for comparison. Inset: the
coefficient $C$ (solid) and the `classical limit'\ $C_{\mathrm{cl}}$
(dashed) vs. doping $x$, in which the experimental data are marked by full
triangles \protect \cite{Phase,NMR,SDWtransition}.}
\label{invchi}
\end{figure}

\emph{Mean-field approximation.---}The spin dynamics is governed by $H_{s}$, where $A_{ij}^{d}$ given in Eq. (\ref{ad}) is only density-dependent and \ can therefore to leading order approximation be treated as a smearing flux of $\pi (1-x)$ per unit cell. The resulting `mean-field' state of $H_{s}$ is Landau quantized with spins being statistically `localized'\ in the cyclotron orbits, in sharp contrast with a degenerate Fermi liquid state or a fully localized classical state at $t\rightarrow 0$. In particular, the characteristic bandwidth vanishes, while $t$ remains relatively large, when all the spinons stay in the degenerate LLLs at low temperatures, rendering a systematic `scaling' behavior as will be shown below \cite{remark}.

\emph{Curie-Weiss uniform susceptibility.---}One peculiar property is exhibited by the uniform spin susceptibility, $\chi _{u}={\frac{2\mu _{\mathrm{B}}^{2}\beta }{N}}\sum_{m}(n_{m}+1)n_{m}$\ where $\beta ^{-1}=k_{\mathrm{B}}T$ and $n_{m}=1/[e^{\beta \left( E_{m}-\mu _{s}\right) }-1]$ is the Bose distribution for spinons at state $m$ with energy $E_{m}$\ obtained by the aforementioned mean-field solution based on $H_{s}$ ($\mu _{s}$ is the chemical potential). As clearly illustrated in Fig. \ref{invchi} by $1/\chi _{u}$, it follows a Curie-Weiss law $\chi _{u}=\frac{C}{T+\Theta }$
with $\Theta =0$ (full squares). A finite Weiss term $\Theta \simeq 3(1-x)J/k_{\mathrm{B}}$ (full circles) is generated by including an RPA
correction from $H_{J}$ (by fitting with the experimental $\Theta $ \cite{Thermopower}, we estimate $J\simeq 70$ $\mathrm{K}$). The coefficient $C$
is $x$-dependent, and at $T\rightarrow 0$ one finds $C=2(1-x)\mu _{\mathrm{B}}^{2}/k_{\mathrm{B}}$. As shown in the inset of Fig. \ref{invchi}, $C$ is in
quantitative agreement with the experimental data \cite{Phase,NMR,SDWtransition} in the Curie-Weiss regime of the cobaltates, which is independent of other parameters in the model, like $t$ and $J$.

By contrast, a Pauli-like susceptibility is expected for the Fermi-liquid state (dotted line in Fig. \ref{invchi}, obtained with a bare $t=-0.1$\textrm{eV }\cite{Singh}). It is particularly instructive to compare this with the classical limit\ of the t-J model at $t\rightarrow 0$ ($J=0)$, where all the electrons are fully localized as $\left( 1-x\right) N$ free
moments, contributing to a Curie's law $\chi _{u}^{\mathrm{cl}}=C_{\mathrm{cl}}/T$ (dashed line in Fig. \ref{invchi}). But one finds $C_{\mathrm{cl}}=C/2$, i.e., only about the \emph{half} of the values of both the experimental and the present theory (cf. the inset of Fig. \ref{invchi}). Clearly the peculiar quantum effect of the present bosonic spinons is responsible for the enhancement of $C$ from $C_{\mathrm{cl}}$ at low $T$ (only at high-$T$ limit, the `classical' $\chi _{u}^{\mathrm{cl}}$ can be recovered as $n_{m}\rightarrow 0$ in $\chi_{u})$.

\begin{figure}[tb]
\includegraphics[angle=0,scale=0.3]{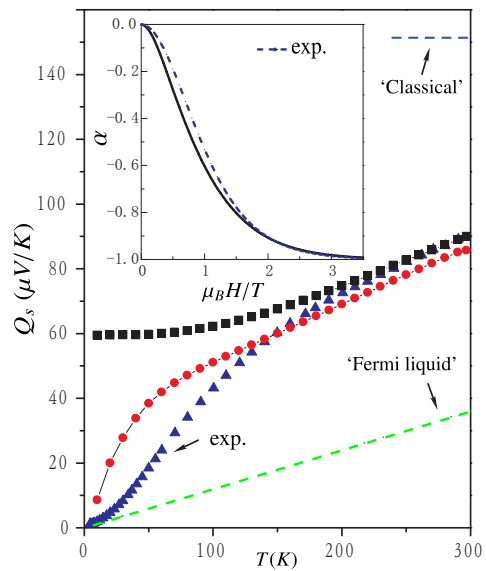}
\caption{Thermopower $Q_{s}$ contributed by the spin part at $x=0.71$. Full
squares: the mean-field solution; Full circles: the effect of doublon
density fluctuations\ is considered (see text). The experimental result
(full triangles) at the same $x$ as well as the Fermi liquid (dotted) and
`classical limit'\ (dashed) results are shown for comparison. Inset: The
scaling curve (solid) for $\protect \alpha $ defined in Eq. (\protect \ref{alpha}) vs. $H/T$ and the spin-entropy of free moments (dashed), which well accounts for the experiment measurement \protect \cite{Thermopower}. }
\label{thermo}
\end{figure}

\emph{Thermopower.---}In Fig. \ref{thermo} the thermopowers predicted by the Fermi-liquid state as well as the `classical limit'\ of the t-J model are shown by dotted and dashed curves, respectively. In the latter case, with
the bandwidth vanishing, the thermopower reduces to the so-called Heikes formula: $Q_{\mathrm{cl}}=\left(k_{\mathrm{B}}/e\right) \ln \frac{2x}{(1-x)}$ which is proportional to the entropy per electron \cite{Hubbard1,Hubbard2}. Both deviate strongly from the experimental result (full triangles in Fig.\ref{thermo}) in opposite ways.

In the present saddle-point state, the thermopower will satisfy the Ioffe-Larkin combination rule $Q=Q_{d}+Q_{s}$ according to Eqs. (\ref{hd}) and (\ref{hs}), with the spinon contribution $Q_{s}$ dominant over the doublon part $Q_{d}$. Here a `large' $Q_{s}$ originates in the degeneracy of the LLLs. This leads to a Heikes-like formula $Q_{s}=-\mu _{s}/eT,$
associated with the spinon entropy, which follows given that the energy-particle current correlator $S^{12}=0$ at low temperature in the Kubo formula $Q_{s}=-\frac{1}{eT}\frac{S^{12}}{S^{11}}-\frac{\mu _{s}}{eT}$ \cite{note}. Note that the above Heikes-like formula is still valid even when $S^{12}$ becomes finite, for instance when the spinons are excited to the higher Landau levels at high-$T$, because the current-current correlator $S^{11}\rightarrow \infty $ \cite{long version,semion}.

The calculated $Q_{s}$ is presented in Fig. \ref{thermo} (solid-squares). This saturates at low-$T$ to a `universal'\ value $Q_{s0}=\frac{k_{\mathrm{B}}}{e}\ln 2\approx 60$ $\mathrm{\mu V/K}$ due to the artifact that the bandwidth\ of the LLLs remains zero. But a weak fluctuation of the gauge flux of $A_{ij}^{d}$ around the mean-value can easily cause a broadening
lifting the exact LLL degeneracy, resulting in a vanishing $Q_{s}$ at $T\rightarrow 0$. A typical example is indicated by the full circles in Fig. \ref{thermo}, with $A_{ij}^{d}$ simulated by a random flux of $[-0.05\pi,0.05\pi ]$ around the mean flux of $\pi (1-x)$ per unit cell. One finds a quantitative agreement with the experimental measurement by taking $
t_{s}=100 $ K. Note that $Q_{d}$ contributed by doublons is not included here as it is usually much weaker in magnitude with a quite flat $T$-dependent as calculated by the Kubo formula \cite{long version}.

To understand more clearly the origin of $Q_{s}$ from the spin degrees of freedom, one may apply a strong in-plane magnetic field\ $H$ to polarize (freeze) the spins via Zeeman energy, as already accomplished experimentally
\cite{Thermopower}. Denote $Q_{s}(\infty )$ as the contribution from the remaining configurational entropy at $H\rightarrow \infty $ and define
\begin{equation}
\alpha =\frac{Q_{s}(H)-Q_{s}(\infty )}{Q_{s}(0)-Q_{s}(\infty )}-1.
\label{alpha}
\end{equation}
Then $\alpha $ is found to be a well-defined scaling curve of $H/T$, as shown by the solid curve in the inset of Fig. \ref{thermo}, which fits well with the pure spin entropy of free moments: $\left[ \ln [2\cosh (u)]-u\tanh
(u)\right] /\ln 2,$ where $u=\left( 2.2\mu _{\mathrm{B}}H\right) /\left( 2k_{\mathrm{B}}T\right) $ in excellent agreement with the experimental data as given in Ref. \cite{Thermopower}.

\begin{figure}[tb]
\includegraphics[angle=0,width=\columnwidth]{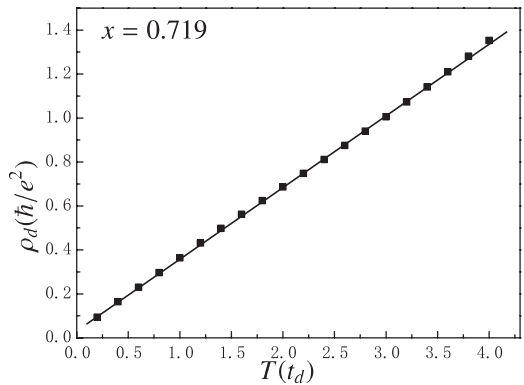}
\caption{The resistivity is contributed by doublons in the presence of
random flux tubes $\pm \protect \pi $ attached to spinons in the LLLs, which
numerically treated as a random distribution of $\pm (1-\protect \delta )%
\protect \pi $ fluxes per unit cell. }
\label{res}
\end{figure}

\emph{Resistivity.---}The resistivity satisfies the Ioffe-Larkin combination
rule $\rho =\rho _{d}+\rho _{s}$ in which $\rho _{s}=0$ due to the
aforementioned \textquotedblleft Meissner effect\textquotedblright \
response \cite{long version,semion} in the spinon subsystem such that the
doublon subsystem will solely contribute to $\rho $. Governed by $H_{d}$,
the doublons will experience the novel scattering generated from the gauge
potential $A_{ij}^{s},$ which describes $\pm \pi $ flux-tubes bound to the
spinons with $\sigma =\pm $. In the low-$T$ regime where the spinons are in
the degenerate Landau orbits, the corresponding flux-tubes seen by the
holons will distribute randomly in space. Based on a numerical calculation
where $A_{ij}^{s}$ is treated as randomly distributed $\pm (1-x)\pi $ fluxes
per unit cell, we obtain a metallic behavior of $\rho _{d}$ with a linear-$T$
dependence over a large temperature regime as shown in Fig. \ref{res},
originating in a scattering rate similar to a case previously studied \cite%
{Gu} in a high-$T$ regime in the context of high-$T_{c}$ cuprates.

\begin{figure}[tb]
\includegraphics[angle=0,width=\columnwidth]{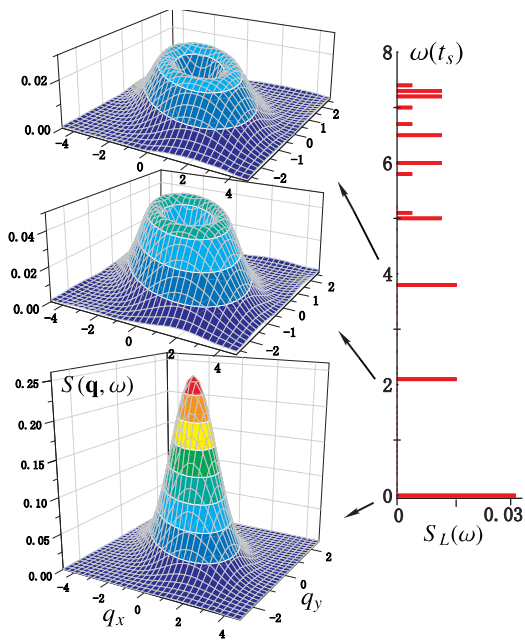}
\caption{Dynamic spin structure factor $S(\mathbf{q},\protect \omega )$
predicted by the mean-field theory at $x=0.75$. Right panel: the Landau
level structure exhibited in the local ($\mathbf{q}$-integrated) dynamic
structure factor $S_{L}(\protect \omega )$; Left panel: the corresponding $%
\mathbf{q}$-dependence of\textbf{\ }$S(\mathbf{q},\protect \omega )$ for the
first three energy levels.}
\label{spindynamics}
\end{figure}

\emph{Predictions and discussion.---}The fractionalized saddle-point state
governed by (\ref{hd}) and (\ref{hs}) has been shown to generically exhibit
a systematic scaling behavior: $\rho \propto T$, $\chi _{u}\propto 1/T,$ a
large thermopower $Q_{s}\sim \frac{k_{\mathrm{B}}}{e}\ln 2$ as well as the
scaling law in $\alpha =\alpha (H/T)$, associated with the peculiar Landau
quantization effect in $H_{s}$. Our theory allows us to make one further
strong prediction: the spin sector Landau levels can be in fact directly
observed by inelastic neutron scattering! The dynamic spin structure factor $%
S(\mathbf{q},\omega )$ can be easily computed assuming the straightforward
Landau quantization and it should look like Fig. \ref{spindynamics}: a tower
of rather narrow non dispersive bands of spin fluctuation (right panel),
with a momentum dependence as indicated in the left panel.

Another interesting consequence of the spin-charge separation is in the
figure of merit, used to quantify the efficiency of a thermoelectric device.
This can be written in the following form
\begin{equation}
ZT\equiv \frac{Q^{2}T}{\rho \kappa }=\frac{(Q_{s}+Q_{d})^{2}/L_{d}}{1+\kappa
_{ph}/\kappa _{d}}  \label{ZT}
\end{equation}%
since the spinons do not contribute to the thermal conductivity and
resistivity ($\kappa _{s}=0$ and $\rho _{s}=0$) \cite{long version}. Then a
large spinon thermopower $Q_{s}$ ($\gg \left \vert Q_{d}\right \vert $) will
play an important role, independent of the Lorenz number of the doublons $%
L_{d}\equiv {\frac{T}{\kappa _{d}\rho _{d}}}$ as well as the doublon ($%
\kappa _{d})$ and phonon ($\kappa _{ph}$) thermal conductivities. We find
\cite{long version} $L_{d}$ to be one order of magnitude smaller than the
Lorenz number of a Fermi liquid. This is in fact consistent with exact
diagonalization result on the t-J model \cite{shastry1}. The combination of
these factors may therefore lead potentially to an exceptionally large $Z$.

Finally, we already alluded to the fact that various factors leading to a
broadening of the Landau levels are ignored in the present mean-field
treatment \cite{remark}. Accordingly, the physics we have discussed here is
associated with an intermediate temperature regime, bounded by the Landau
gap scale at high temperature and the onset of further cooperative phenomena
at low temperatures. In the latter regime the specific heat measurements
have also shown\cite{gamma1,gamma2} an anomalous increase, presumably due to
the afore-discussed entropy enhancement which, however, will be sensitive to
the detailed Landau level broadening.  Eventually the spin degeneracies
associated with the Landau states will be lifted, and this might be the
origin of some new emergent orders at very low temperatures. One infers from
the momentum dependence of $S(\mathbf{q},\omega )$ at the LLLs (Fig. \ref{spindynamics}) that the in-plane ferromagnetic correlation length will increase with $x$, such that a small interlayer superexchange $J_{\perp }$may eventually drive an in-plane ferromagnetic order with antiferromagnetic ordering along different layers, explaining the $A$-type antiferromagnetic order observed by neutron scattering at $x\geq 3/4$ \cite{NIS}. The bosonic
doublons may also condense at a sufficiently low-$T$ to make the system more Fermi-liquid like, reconciling with the Wiedemann-Franz law observed in \textrm{Na}$_{0.7}$\textrm{CoO}$_{2}$ at $T<1$ \textrm{K }\cite{SYLi}.
However, our unconventional `building material' also leaves room for less conventional forms of order, including uncommon charge orders that will be discussed in detail elsewhere.

\emph{acknowledgments---}We acknowledge stimulating discussions with N.P Ong, X.L. Qi, and in
particular Y. Wang who also provided us with unpublished experimental data.
This work was supported by the grants of NSFC and the National Program for
Basic Research of MOST, China, and a Spinoza grant of the Nederlandse
Organisatie voor Wetenschappelijk Onderzoek (NWO).


\begin{thebibliography}{99}
\bibitem{remark} We emphasize that in principle a sharp LLL degeneracy is an
artifact of the mean-field approximation which can be easily lifted, for
instance, by density fluctuations via $A_{ij}^{d}$ intrinsically or
extrinsically (e.g., induced by the disordered distribution of \textrm{Na)}.
Nonetheless the mean-field results is expected to remain a good
approximation to the saddle-point physics at `intermediate' temperatures
higher than the LLL broadening.
\bibitem{note} Here $\mu _{s}$ is defined with taking the LLL energy $%
E_{0}=0 $ such that $\mu _{s}(T=0)=0.$
\end{thebibliography}
\end{document}